# Towards a Science of the Mind
Jerome Feldman
July 10, 2019

## Abstract

The ancient *mind-body problem* continues to be one of deepest mysteries of science and of the human spirit. Despite major advances in many fields, there is still no plausible link between subjective experience and its realization in the body. This paper outlines some of the requirements for a rigorous science of mind (SoM) - key ideas include *scientific realism*, *agnostic mysterianism*, careful attention to language, and a focus on concrete questions and results. A core suggestion is to focus effort on the (still mysterious) mapping from bodily activity to subjective experience.

## The Scientific Reality of Mind

The mystery of the mind is now often called "The Hard Problem". The popular 2015 Tom Stoppard play by that name is available on the web. The core of the mind-body mystery has always rested on our phenomenal 1$^{st}$ person feelings of subjective experience (SE). We now know more than ever about the structure and function of the body and brain, but there is no comparable science of SE in an embodied mind.

The philosopher, David Chalmers, who has thought as much as anyone about the hard problem (35) says:

*"I do not claim that idealism is plausible. No position on the mind–body problem is plausible. Materialism is implausible. Dualism is implausible. Idealism is implausible. Neutral monism is implausible. None-of-the-above is implausible."*

On the other hand, research often shows a direct correlation between the subjective experience of color, pain, etc. and measurable bodily function. Even in these direct situations, the neural (or other) physical realization of SE remains unknown and this mystery is called the *explanatory gap* https://plato.stanford.edu/entries/qualia/. A crucial first step towards a SoM is to acknowledge that there is a problem and I am suggesting a position of *agnostic mysterianism* that is plausible. However, it is currently conventional, almost mandatory in some fields, to assert (without evidence) that there is no problem. This quote from the distinguished clinician and neuroscientist Michael Gazzaniga is typical:

"*Instead of an immaterial mind floating around within each of us, modern science has moved the mind into the brain and made it very physical*" (10, p.28).

Taken literally, such statements are proposed general scientific laws (https://explorable.com/falsifiability) of the form: There is a scientific concept (class) C such that for any c in C, the property p(c) holds. In the case above, the assertion is that every subjective experience has a neural cause. Following Popper, any such general claim can be falsified by an established counter-example. Several such counter examples are discussed in this paper and in (3). This article is my attempt to suggest a set of ideas and attitudes that could form a common basis for a scientific approach to some mysteries of the mind. It assumes that there is a deep problem to be addressed.

There is no evidence for SE in the absence of neural activity so there must be some profound connection between brain and mind. However, there are also many cases where the mapping between SE and any corresponding embodied substrate remain mysterious. For example, natural language input can evoke a wide range of emotional and other subjective experiences. At a more basic level, the perception literature contains many examples of constancies and "illusions" where SE is inconsistent with current or proposed theories of neural computation (3).

One of the most common unexplained examples is the SE of continuous motion from sequences of still images in films, videos, and normal vision. This situation is nicely summarized in the following 3/14/19 personal communication from Ken Nakayama:

"*Despite the fact that this class of Gestalt psychology type demonstrations have been known for over 100 years, the actual physiological basis of any of them is unknown. In fact, it is something that I covered in my large and lengthy review (34) of image motion processing in 1985. I wager that there is no known satisfying physiological understanding of perceived motion experience most generally. This despite the fact that there are hundreds of papers on the physiology and psychophysics of motion.*"

The discrepancy between our subjective experience, sensory input, and the current understanding of brain function is also discussed in detail in Dehaene (26, pp142 ff.). If we acknowledge the mystery, the question arises of how it could be addressed scientifically. Any science of mind is necessarily committed to *scientific realism*. As is often the case, Rebecca Goldstein puts it best in her contribution to the book: What Scientific Term or Concept Ought to be More Widely Known? (1)

"*Scientific realism is the view that science expands upon—and sometimes radically confutes—the view of the world that we gain by means of our sense organs. Scientific theories, according to this view, extend our grasp of reality beyond what we can see and touch, pulling the curtain of our corporeal limitations aside to reveal the existence of whole orders of unobserved and perhaps unobservable things, hypothesized in order to explain observations and having their reference fixed by the laws governing their behavior. In order for theories to be true (or at any rate, approximations of the truth) these things must actually exist. Scientific theories are ontologically committed.*"

A paradigm example is the Greek postulation of the atom as the underlying constituent of matter. The contemporary proposed 18 sub-atomic particles is another case. Of course, scientific realism, like all of science, is approximate and subject to revision. This entails the fact that hypothesized scientific concepts will sometimes be revealed as mistakes. Famous cases include the ether, phlogiston, vitalism, and empty space. I will suggest below several current concepts that are also poorly defined and should be added to the list of scientifically un-useful terms.

Any development of a SoM depends on the realization that the *mind-body-world problem* is currently a *mystery*. The perception literature already contains a wide range of touchstone challenge problems (2, 3, 34) for any proposed reductionist explanation of the relation between the human mind, our bodies and brains, and the physical world, including Figure 1, which will be discussed below. Any SoM must also respect the resource (time, space, etc.) constraints on adaptive behavior.

**Technical Vocabulary for a SoM**

*But I've gotta use words when I talk to you.*" Sweeney in T.S. Eliot's Sweeney Agonistes.

Sweeney's lament here was that words could not express the full complexity of his feelings and thoughts. Terminology alone will not solve any hard scientific problems, but confusing terminology can be a significant barrier to progress. A current example is the recent spurt of papers arguing that the term "consciousness" should be applied to various different animals. Such preemptive claims only hinder ongoing progress on the abilities and experiences of animals.

The contested usage of technical terms is endemic in discussions of two of the most profound scientific mysteries: the mind-body problem and the ultimate nature of physical reality (28). Even the word "mystery" has a range of meanings. A standard dictionary entry is - *Something that is difficult or impossible to understand or explain*. Scientific usage is even broader, ranging from minor uncertainty to inherently unapproachable. I will follow the common idea that a scientific mystery is a phenomenon for which there is currently no known approach to a solution.
.
There are many scientific mysteries, but few are as confounded by language issues as the classical mind-body problem. For one thing, language is itself a fundamental, possibly essential, feature of human thought. The standard meaning of "mind" in the mind-body problem, which we will follow, is the aware personal subjective experience (18) of individuals. Investigation of related phenomena including subliminal perception, dreams, and imagination could be helpful. When we include the additional mystery of how the human mind and body evolved to thrive in the physical and social world, we get the mind-body-world problem.

The mind-body-world problem is also of continuing interest in Philosophy (35), where the meaning of technical terms is even more important. The classic book (25) by Hillary Putnam contains a detailed analysis of fundamental lexical difficulties in the philosophy of mind.

Even the naming of the problem can have a profound effect on research. The ancient Greeks had no knowledge of brain function and talked about the mind-**body** relationship. (Thales: Νοῦς ὑγιὴς ἐν σώματι ὑγιεῖ   A healthy mind in a healthy body.) Any current presentation of the mind-**brain** problem entails a presupposition that the mind is a solely function of the brain. In fact, we all know that hormones, microbes, health, and many other aspects of the body can have profound effects on subjective experience. More confusion arises when the term "mental" is used to refer to the brain and to thinking in general, as in mental function, etc.

Historically, Democritus built a theory of perception explicitly based on his famous idea of atomic structure.
"*Democritus' theory of perception* depends on the claim that eidôla or images, thin layers of atoms, are constantly sloughed off from the surfaces of macroscopic bodies and carried through the air … It is the impact of these on our sense organs that enables us to perceive."
https://plato.stanford.edu/entries/democritus/

This was probably the first enunciation of the *mind-body-world problem* that examines the relations between the physical world, our bodies (and brains), and first person subjective experience. From a contemporary view, human bodies and minds evolved to be adaptive in the physical and social world, so the mysteries of the mind and of physical reality are entangled (28).

Without a commitment to the scientific reality of "mind" as a subject of scientific inquiry, there can be no science of the mind (SoM). The very idea of a SoM has been problematic. There are some profound reasons for this, but much of the difficulty arises from the lack of core technical vocabulary. The introduction to the ambitious book: *The Neuroscience of Emotion: A New Synthesis* (24) nicely expresses the importance of technical language for science:
"*A comprehensive science of emotion also needs to connect with all domains of science that are relevant to emotion: it needs to connect with psychology and with neurobiology. Doing this requires a consistent terminology that makes principled distinctions, and that allows clear operationalization of the different concepts that a science of emotion will use.*"

The emotion community has made a significant terminological advance by (largely) agreeing to refer to the SE aspects of embodied *emotions* as "f*eelings*". This has enabled the authors of (24) to state straightforwardly (p12 ff.) that the book will focus on the scientifically tractable issue of emotions and defer the mind-body problem of feelings.

There are many other disciplines and practices concerned with the human mind and the foundations for scientific study are not always clear. For example, a statement that a spider or a robot has a mind could be a scientific or a definitional assertion. Similarly, when philosophers postulate "the extended mind",
*https://en.wikipedia.org/wiki/Extended_mind_thesis*, this could be a profound ontological claim or just the obvious fact human minds depend upon cultural artifacts and communities.

Despite these challenges, there is interesting and productive work in the SoM (2), but no coherent program. One requisite is a small core vocabulary (24) of technical terms for a unified SoM. Almost all of the terms proposed here are common words with standard meanings. To begin, *Science* is public exploration of phenomena, involving theory, modeling, experiment, and replication. As above, *Mind* canonically refers to the private mental activity of humans. These two definitions already highlight one of the challenges for any SoM – how can there be a public science of private experience. In addition, there are many extended and metaphorical uses of terms like *mind* and often confusion about their intended meaning.

A *scientific problem* or *mystery* is a phenomenon for which there is currently no plausible explanation. A related source of mystery is an *inconsistency* between two or more theories of the same phenomena. A previous article (3) made the case that the mind-body-world problem is *inconsistent* with current neuroscience and computational theory. Such inconsistencies often lead to scientific revolutions. Much of the historical success of science can be traced to concerted effort on mysteries. One of the best-known cases is the fact that Rutherford's planetary model of the atom entailed that electrons orbiting around the atomic nucleus would radiate energy and eventually crash into the nucleus. This was one of a number of deep inconsistencies leading to the development of quantum theory.

A recent important inconsistency in Cognitive Science is the "Word Superiority Effect" (23). A wide range of experiments established that people were faster and more accurate in recognizing the letter A in context, e.g., CAT, than the same letter alone. These results conflict with the naïve assumption that more input should require more processing. This was one of the inconsistencies resulting in the paradigm shift to massively parallel (connectionist) models of brain function   https://en.wikipedia.org/wiki/Connectionism.

My most radical proposal here is that we focus on the particular general mystery of the *mapping* from systemic bodily activity to SE. This is clearly part of the mind-body-world problem and does not require any philosophical or other non-scientific considerations. Following the principle of Scientific Realism above, I propose a name for this postulated mapping. My current suggestion is the Greek letter *X*, named "Chi" and pronounced "kai". For reasons beyond the scope of this paper, any talk of an inverse mapping would put us in a deep theological morass - https://en.wikipedia.org/wiki/Downward_causation. The important subcase of perceived continuous motion is often called the Phi phenomenon (34) and is thus part of the *X* mystery. The key goal here is to recast a core problem of mind to a standard scientific mapping question that should be approachable by existing methods. The proof that existing methods do not suffice (3) exhibits the kind of inconsistency that is often productive in science.

I also suggest that we use the terms *evolution* and *fitness* in the conventional manner as they apply to all living things. A companion paper (ref) will discuss the crucial role of evolution for any concept of mind. There is no way for an organism or a scientist to calculate present fitness, since it depends on the future. We will use a somewhat new term *actionability* (4) to label an organism's internal estimate of the expected fitness of its potential actions in the current *situation*. Actionability is a consequence of autopoesis https://en.wikipedia.org/wiki/Autopoiesis. Any living system will include an internal mechanism for choosing actions.

In simple cases like the amoeba or human reflexes, actionability reduces to immediate *action*. Actionability is an extension of Gibson's perceptual *affordances*, https://en.wikipedia.org/wiki/Affordance that adds quantitative, active, situational, goal-directed perception and reasoning. One major function of a mind involves comparing actionabilities, sometimes by mental *simulation* (4), before committing to an action. It is not feasible to pre-compute actionability for novel contexts so active simulation is needed.

The main additional requirement for a SoM is *scientific realism,* described above. As scientists, we must start with the assumption that the mind, body, and the external world are all real in the sense that they are subjects of study. Many other practices and belief systems (including reductionist physicalism) deny the reality of one or more of these. Any such denial is the most serious barrier to a SoM. Even without denying the reality of the mind, it is popular to assume that there is no mystery, that the mind is manifestly the activity of the brain, and that all will be revealed by routine current science, so there is no need for a SoM (9). Stanislas Dehaene, an outstanding experimentalist, says, "*If you had any lingering doubts that your mental life arises entirely from the activity of the brain, these examples should lift them*" (26, p.153).

There are many well-known scientific challenges to such faith-based reductionist theories of mind. The figure below is one of the simplest.

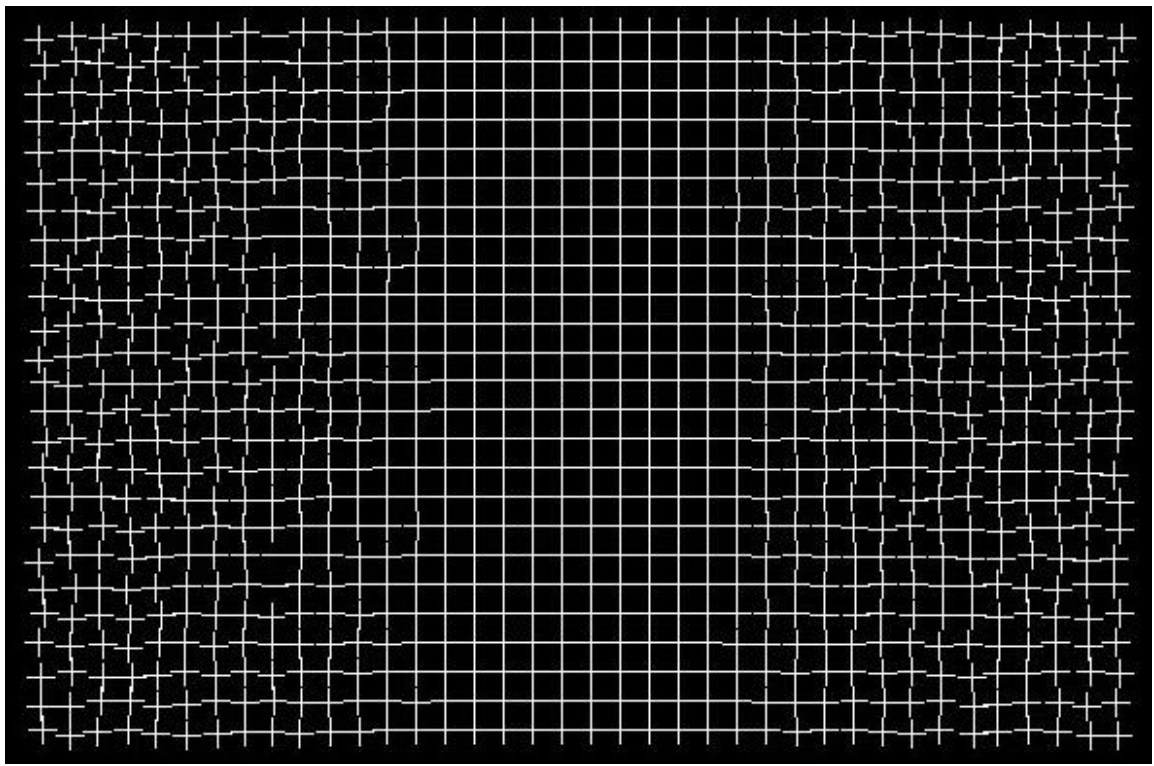

Figure 1. A Perceptual Mystery

Focus on the center of the image for about a minute and the grid will appear uniform. There is nothing in the brain as we know it that could encode the large regular grid that we experience.

Some other basic inconsistencies among neural computation theory, brain function and visual experience are described in (3). There are also many other mysteries of SE. For the most part these can be of two different types:
1) Emotions and Feelings; prototypically pain
2) Perceptions (including proprioception); prototypically color

These are not mutually exclusive and are generally treated as a single mystery. Both seem to fall under the *X* mapping mystery. One difference is that feelings like pain are sometimes attributed to animals with primitive sensing but who lack perception.

Several more or less equivalent terms are employed to describe the *subjective experience* (SE) associated with the mind, but SE appears to be the most focused and least ambiguous. For the core vocabulary, we will consider the following as sharing meaning: *Phenomenology, Subjectivity, 1st person Experience*. Single terms like *experience* and *feeling* are sometimes employed, but they have numerous other uses. The term *Qualia* is also used, but it has too many versions for our purposes. Philosophers often talk about the 1st person experience of "what's it like" in this context. Since all of these terms (like mind) refer by default to the human mind, we need another term to denote mind-like behavior in (much) simpler animals.

No one seriously argues against the idea that (many) mammals have experiences (27) that share some of the properties of human emotions like pain, fear, anger, joy, etc. It is also clear that there are important differences, at least in that only people can talk about such experiences. If we assume that subjective experience has evolved then, there must be evolutionary *precursors* that also need to be studied (5, 6). Evolutionary considerations will be discussed in a forthcoming paper (ref).

In addition, there is currently no good term to label any postulated mind-like capacity of robots and this leads to considerable difficulty. We will probably not be able to define the robot equivalent of SE until we better understand animal SE and its *precursors* (7). Using the term "precursors" can replace the intense arguments on whether various animal experiences should be labeled "mind" or "consciousness".

One useful distinction is the recent C0, C1, and C2 definitions in Dehaene et al. (9). They define C0 as unconscious neural activity that is not experienced, C1 as activity that is accessible for computation and report, and C2 as self-monitoring of C1. The discussion in (9) focuses on some simple information processing functions of C1 and C2, but the classification is much more general. A major advantage of this proposal is that it provides a label (C0) for the vast range of behavior in the absence of awareness. The functions of C0, C1, and C2 are not mutually exclusive and often overlap, for example, when you stumble while walking.

Human C0 behavior ranges from reflexes and other intrinsic circuits to homeostasis and language understanding. In fact, much human activity (e.g. driving) requires automatization to C0 and cannot be done mindfully as C1. In this terminology, SE falls within C1. A complication is that automatized behavior (C0) can give rise to SE. For example, unconscious interoception systems evoke SE of hunger, fatigue, etc. Emotions are normally automatic, but definitely involve SE ((24, p. 83)).

There is general agreement that subjective experience (SE) is an essential aspect of the mind and of any notion of consciousness - we will focus on SE unless specifically stated otherwise. It is quite possible that some animals have C1 but not C2; C2 does seem necessary for language.

**Words to avoid.**

There are also terms that should be explicitly excluded from the SoM core vocabulary – words such as consciousness, dualism, illusion, instinct, and technical terms from other domains like – AI, quantum, recursion, supervenience, God, etc.

The term "illusion" has two mutually inconsistent definitions and is often used without specifying which is intended. The word is sometimes used to describe a perception that is inconsistent with external reality and at other times to describe an experience that is instead inconsistent with the neural representation, even when the subjective experience is **more** like external reality (3). Further confusion arises when "illusion" is used metaphorically, as in the postulated "illusion of Free Will". At least most Westerners agree that we all act as if we had Free Will in everyday life, even determinists who deny that they have this capability. There does not seem to be an agreed upon definition of "illusion" that supports its use in a serious discussion of the mind.

The term "consciousness" has dozens of meanings, and there is currently no standard scientific definition. George Miller (8) suggested in 1962 that use of the term consciousness should be suspended.

"*Consciousness is a word worn smooth by a million tongues. Depending on the figure of speech chosen it is a state of being, a substance, a process, a place, an epiphenomenon, an emergent aspect of matter, or the only true reality.*"
Much of the work discussed here does explicitly employ the term consciousness and I will follow their usage with the understanding that subjective experience is the issue of current concern.

Similarly, the word "qualia" is a highly contested concept sometimes defined as identical to neural activity, sometimes epiphenomenal, and sometimes as completely unconnected to the brain. My preferred term, *subjective experience (SE),* covers both neural activity and the currently mysterious accompanying private experience and it assumes that there are connections to be discovered. The unknown mapping from neural activity to SE is what I am calling *X*.

I also suggest that the term *instinct* has no technical value in a SoM, especially when it is proposed as a scientific answer to the mind-body problem as in Gazzaniga's excellent book "The Consciousness Instinct" (10). An instinct is a behavior that depends, at least in part, on specific genetic factors rather than only on learning. There is certainly such a substrate for the human mind, but applying the label "instinct" does nothing to illuminate the mystery of how it works.

A major linguistic source of confusion is the dichotomy *monism/dualism*. This can lead to the assertion that either our current scientific knowledge is adequate or if not we must invoke some non-material ontological forces. The most toxic term for a SoM is *dualism*. This word is invoked to suggest that any questioning of current reductionist materialism necessarily invokes some immaterial spirit that is outside of science. The quote above from Michael Gazzaniga is typical and there are many others. Science does not work in this restricted way. The history of science is replete with the discovery of novel theoretical concepts having the most profound scientific and practical consequences. The history of Medicine presents a paradigm example.

In summary, any effective SoM must honor the postulated scientific reality of the mind, the body, and the physical world. It must also acknowledge that there are currently scientific mysteries, including the mind-body-world problem. We should not assume that the "causal closure of the physical" implies that current science is complete. Given the great historical success of scientific inquiry, we should assume that some, but not necessarily all, aspects of these mysteries can and will be explained in this century. The position that best summarizes these constraints is *agnostic mysterianism* (3, 11). Agnostic mysterianism is neutral with respect to hypothesizing one or more aspects of mind that are not currently reducible. Importantly, this approach rules out postulating particular non-physical influences (idealism). We do require the testing of any SoM proposal on touchstone problems like Figure 1, motion perception, the binding problem, and the stable visual world (3).

**Beyond Words**

Science is only one of many approaches to coping with the at least currently provide all the answers. In fact, most people do not directly rely on science for any of their important beliefs and actions. Nevertheless, Science, and more generally intellectual activity, continues to have profound impacts on the world and our understanding of reality, including our minds.

This article is an attempt to suggest a set of ideas that could form a common basis for a scientific approach to some mysteries of the mind. Obviously, this requires a minimal shared understanding of various terms, as proposed above. Actually doing the science will entail making additional distinctions, as always. For concreteness, I will follow the literature and mainly discuss visual perception, but will also briefly consider subjective aspects of other perception and actions.

**Clinical Findings, Concreteness, and Computation**

Very few of the myriad theories of the mind, consciousness, etc. take into account the relevant rich clinical findings. Damasio (12) and Gazzaniga (10) are significant exceptions and should be required reading. One major clinical outcome is that some aspects of SE and mind depend entirely on primitive brainstem structures found in at least all mammals. Among other things, this moots the arguments about whether consciousness is based in the front or the back of the cerebral cortex. These results also challenge the position (9) that meta-cognition (C2) is required for subjective experience (C1). In addition, these clinical findings add to the importance of studying the precursors of mind in other animals (6, 7).

Even the restricted notion of Mind defined here might be too broad for an initial integrated effort on a Science of Mind. There are several advantages to starting with *concrete* cases of mental activity, like perception and action, which have measureable correlations with the external world. This involves reduced emphasis on interesting questions like emotions, free will, the sense of self, etc. An obvious benefit of this concrete approach is that we know that animals also depend on effective perception and action. In addition, the behavior and neural structure of some mammals is similar to that of humans and is much better understood.

Many of the historic mind-body issues, like the unity of perception, aka the binding problem (16), and the internal model of the physical world (5) are also present in animal behavior. Most of the intellectual effort on SE has focused on vision, but there are also classic mind-body issues in speech, body awareness, and motor action. https://plato.stanford.edu/entries/qualia/   . This site also discusses the "explanatory gap" and several conflicting notions of the term qualia.

 An important advantage of concrete functions of mind is that they provide *touchstone* tests of theories and models of mental behavior. In fact, there are demonstrations that the standard model of neural computation (14) is inconsistent with subjective experience including the famous examples of the binding problem and of the stable visual world. A proof of this inconsistency was published as "Mysteries of Visual Experience" (3). That article starts with an experiential demonstration of the stable world "illusion" related to Figure 1 above. It then proves that the basic facts about the structure and behavior of the visual system according to the standard neural theory of computation are *inconsistent* with the experience. It also shows that no known alternative theory of brain computation can explain such mysteries.

There are multiple active communities exploring a wide range of alternative brain realizations of the mysteries of the mind, but none that solve the touchstone problems above. The core technical constraint is that all known fast, non-local communication in the brain (and body) is mediated by neural spike signaling (3). Therefore, any proposed sub-neural computational mechanisms (glia, synapses, microtubules, quantum effects, etc.) are restricted to a single neuron and can only transmit simple signals.

In addition, by focusing on concrete examples we can calibrate the computational resources (time space, etc.) required for proposed non-neural realizations of the touchstone examples (3). There is currently no plausible computational alternative to the standard model (14), although such an alternative is not precluded and appears to be required.

This focus on concrete mind-body experience has yet another advantage – we can study precursors of these mental activities in simpler animals. If we assume that the human mind is a natural trait, we should study how it might have evolved (5, 6, 7). This will be discussed in detail in a companion paper (TBD). Looking ahead, many animals have the subconscious (C0) functional equivalent of concrete experiences such as sensory integration and spatial awareness. Moreover, even current robots exhibit some of these abilities. However, the core mystery of SE (mind) remains unexplained.

**Science as Demystification**

*'The most beautiful and profound experience is the feeling of mystery. It underlies religion as well as all deeper aspirations in art and science.'*  Einstein

A core mission of science is attempting to explain the mysteries of nature. The history of science is largely a saga of increasingly sound theories of the physical and social world. There is broad agreement that the nature of the mind is one of the deepest current mysteries and one might hope that science will help demystify it. In fact, several ancient mysteries of the mind have been largely reduced to routine science within our lifetime.

One interesting case is synesthesia (15), a perceptual experience in which stimuli presented through one modality spontaneously evoke sensations in an unrelated modality. The most common form, seeing/hearing numbers or letters as having specific colors, was well known to the Greeks. Many eminent scientists have studied the phenomenon including Goethe, Locke, and Newton.

The father of psychophysics, [Gustav Fechner](Gustav Fechner) reported on a first empirical survey of colored letter experience among 73 synesthetes in 1871. There followed several decades of empirical work, interrupted by the dominance of behaviorist strictures against subjective experience for a period around 1920-1965. The results included the fact that synesthesia had a genetic component, but not much more. Active research resumed in the1980s as a consequence of the general cognitive revolution and this effort established the reality of synesthesia, but the mechanisms were still unknown.

The scientific understanding of synesthesia has advanced through a wide range of theoretical and experimental efforts. The empirical genetic link has given rise to a rich collection of detailed studies of the individual genes involved in various instances of the dozens of known synesthesias. One early theoretical hypothesis was that some extraneous neural connections survived pruning in early development; there is now considerable support for this and it has led to discussion of why this "error" has evolutionary value (13). The full panoply of modern imaging methodologies is also being applied and these supply strong support for the subjective reports. Looking ahead, the combination of physiological measurements with individual subjective reports could provide a solid foundation for a SoM studies.

By now, the science of synesthesia is fully developed, with its own structure of journals, conferences, web presence, etc. and is recognized as a fruitful domain for the study of the genetics, development, neuroscience, behavior, and subjective experience in general (15). This is a paradigm example of demystification in the science of mind. Interestingly, the more exotic synesthesia phenomenon is now much better understood than the more basic (and ancient) problem of how the brain binds multiple features (16).

Because such core mysteries persist, there are active efforts to somehow reformulate the mind-body problems. One of the currently most popular paradigms is often called the 4E (embodied, embedded, extended and enacted cognition). A good overview of this appears in the book (31) and especially in the summary chapter (32). None of this work has yet been made sufficiently concrete to address the standard mind-body mysteries or even the myriad findings of conventional neuroscience. In fact, the 4E enterprise is sometimes viewed as a "philosophy of nature" suggesting interactions beyond the scope of any science (33). From this perspective, the 4E movement has a great deal in common with the much earlier Buddhist" 5 Aggregates of the Mind " https://en.wikipedia.org/wiki/Skandha .

Of course, the mind also entails spiritual and other beliefs where science has not been notably effective. One currently popular reaction to this fact is the movement to defer any direct science of mind and focus on measurable "correlates" of mental experience. This is most developed in the large *Consciousness* community as NCC, the Neural Correlates of Consciousness –

"*Discovering and characterizing neural correlates does not offer a theory of consciousness that can explain how particular systems experience anything at all, or how and why they are associated with consciousness, the so-called [hard problem of consciousness](#) but understanding the NCC may be a step toward such a theory. Most neurobiologists assume that the variables giving rise to consciousness are to be found at the neuronal level, governed by classical physics, though a few scholars have proposed theories of [quantum consciousness](#) based on [quantum mechanics](#)*"   https://en.wikipedia.org/wiki/Neural_correlates_of_consciousness

This NCC effort is perfectly compatible with the more focused SoM suggested here and might well produce useful insights. However, it is also compatible with the belief that the mind is an epiphenomenon and of no direct scientific interest. The standard mantra is "the mind is what the brain does" with no details provided. NCC contains no general acknowledgement of the inconsistency (3) of the standard model of neural computation (14) with our subjective experience.

Given these formidable challenges, I suggest that the SoM effort focus on systematically demystifying specific mysteries as, in the case of synesthesia. Some promising opportunities are discussed below.

**Some opportunities and experiments**

Many disciplines are potentially relevant to a SoM, but direct comparison of the mind and bodily activity seems to be the most promising approach. As mentioned above, there are several research efforts focused on experimental SoM results. In the fall of 2017, Stan Klein and I ran an interdisciplinary UC Berkeley seminar course that explored "Science and Subjectivity". The lectures and background material from that course is available as   http://rctn.org/wiki/VS298:_Subjectivity  and much of the material below follows this.  A particularly focused book is (2).

The findings described earlier in this paper such as motion perception, the stable world, and binding problems could yield concrete *touchstone* problems for proposed theories of representation, computation, and communication in the brain. The broken grid example of Figure 1 is one simple case. Motion perception, the binding problem and the illusion of a detailed stable visual scene (3) are omnipresent in daily experience, are functionally necessary, and have clear informational requirements. We could ask proponents of speculative brain models how their theory could account for these concrete phenomena. That is, assume your theory is true and show how it helps explain these (or other) touchstone tasks. I have done this informally with several leading proponents of various models and have never heard even a vague claim of adequacy.

Community acceptance of some such touchstone tasks could sharpen the discussion of information processing in the brain. Of course, the deep mind-body problem remains a mystery, but we could require that proposed models of brain function address some of the concrete touchstone problems.

One institutional problem is a separation of research goals into careful experimental work on tractable problems and unrelated speculation on the deep mysteries of the mind. We need research on the omnipresent, but currently unexplained, everyday (visual) experiences. As mentioned above, there is a large range of experiments that explore various aspects of the binding problem (16) and the stable world illusion (17). None of these has yet helped with the core mysteries. However, some recent work has illuminated the boundary between the known and the unknown in related areas. Such examples might suggest ways of approaching the problems raised in this paper.

Any scientific effort on the mind-body problem must deal with the reality that the general problem is currently intractable. The main thrust of this paper is to suggest abstracting out the (still mysterious) mapping from systemic neural activity to subjective experience, for which I am suggesting the name *X* or chi. We have known since the work of Penfield, https://en.wikipedia.org/wiki/Wilder_Penfield , that direct micro-stimulation, and more recently Transcortical Magnetic Stimulation (TMS), can evoke certain subjective experiences.  We do not know any plausible realization of the general *X* relation, but it is still possible to gain important insights into the connection between brain and mind, some of which are outlined here. Following the standard mathematical uses of the term "modulo", we can say that such results explore the mind-body relation modulo *X*. That is, there are systematic regularities linking subjective experience to bodily function, although the fundamental linking mechanism *X* remains unknown. Several specific examples are discussed below and more are available from the 2017 course web site http://rctn.org/wiki/VS298:_Subjectivity .

Cohen et al (18) explore the relation between awareness (consciousness) and attention. The mechanisms of (visual) attention are extensively studied and relatively well understood. An important aspect of attention is saccades, which are often triggered without awareness. On the other hand, the fact that most saccades are not noticed shows that awareness does require attention. Thus, awareness requires attention, but attention does not entail awareness. There is also related work on *change blindness* (22), where major changes between consecutive frames go unnoticed when attention is focused elsewhere.

Another visual mystery that has been somewhat demystified is *postdiction*, where a subsequent stimulus seems to causally affect the percept of an earlier input. For example, a flashed letter can be masked by a somewhat (~100-200 ms) later image in a nearby position. Apparent motion (34) is also postdiction since the trajectory can not be computed until after the second flash. Shinsuke Shimojo has done extensive work on postdiction, recently summarized in (19). In one experiment, an artificial scotoma induced by TMS (Transcortical Magnetic Simulation) was filled in with the color shown later.  The paper describes four compatible plausible neural substrate theories for rapid postdiction and considers memory effects like "hindsight bias" that makes events seem more predictable after the fact. This analysis is then extended to address the well-known controversy concerning action before awareness and its relation to the mystery of free will. The article suggests that the "sense of agency" is another instance of postdiction, resulting in an illusion of action before awareness. This does not solve the mystery of free will versus determinism (20), but does greatly sharpen the question.

An impressive recent result is the work of Chang and Tsao on the neural code for faces (30) in macaque cortex. They found that there were 25 key shape features and another 25 for general appearance. The research involves both single unit and FMRI recording of monkeys viewing 2000 face images for which all 50 of the key features that had been worked out, followed by extensive data analysis and modeling. The macaque encoding involves cells in intermediate visual areas that respond to specific features and neurons in higher areas of Inferotemporal cortex that are sensitive to general appearance differences. As always, the $X$ question how all this is bound into the SE of seeing a face remains open.

Uji et al. (36) present a very promising demonstration of dissociable neural activity associated with the qualitative (SE) impression of stereopsis in the absence of processing binocular disparities. They measured EEG activity while subjects viewed pictorial images of 2D and 3D geometric forms under four different viewing conditions (Binocular, Monocular, Binocular aperture, Monocular aperture). Only the viewing condition predicted to generate the strongest subjective impression of stereopsis (monocular aperture) revealed significantly elevated gamma synchronization. These findings suggest dissociable neural processes specific to the qualitative impression of stereopsis SE as distinguished from disparity processing. They are also preparing an fMRI study of this effect. More findings like this would be very helpful.

Perhaps the most directly relevant research is that of the von der Heydt lab (21) on ambiguous figure-ground scenes like the duck/rabbit and face/vase images and the Necker cube. Employing a wide range of behavioral, neural, and computational findings they have established that *border ownership* is the basis for these and a number of related phenomena, including object identification. That is, in these ambiguous examples, various separating edges are assigned to one percept or the other. The paper also describes in detail a plausible neural substrate. They found convincing evidence for the requirement of top-down input in early visual brain areas V1 and V2. Border ownership was also shown to be central to *filling-in*, as with the blind spot. In addition, the effects of border ownership were shown to extend across saccades and to affect attention. The finding of increased spike synchrony between neurons whose border ownership preferences are consistent with the stimulating object, even when the neurons are widely separated in the cortex, is strong evidence for feedback grouping circuits. This is relevant to aspects of the binding problem (16). Neural implementation of border ownership has recently been confirmed and extended by Hesse and Tsao (30).

More generally, Simons and Rensink (22) have a deep and thoughtful discussion of the prospects and perils of scientific research on problems at the boundary of subjective experience. Their story on the developments around the phenomenon of "change blindness" suggests guidelines for the kinds of work suggested here.

**Conclusions**

Across many fields, it has been productive to ignore the hard problem and focus on tractable issues and their applications and consequences (e.g. 5, 6, 9, 14, 18, 24, 26). There are also several active efforts to find a general solution to the mind-body-world problem, consciousness, etc. or to promote one's favorite solution. My suggestion for a Science of Mind (SoM) is more modest.

It begins with acknowledging that subjective experience is central to mental life and that the scientific relation of this (mind) to its embodiment is currently a mystery. The next step is to follow the tradition of "scientific realism" and to establish "mind" as an important research subject. This entails a technical vocabulary and eschewing the scientific use of undefined terms.

The adoption of an "agnostic mysterianism" stance recognizes current mysteries, but also that science has progressively demystified many deep questions, often leading to major advances in knowledge and the quality of life. It also suggests that it is premature to entertain permanent mysteries or supernatural forces. A major proposal is that we focus on the particular sub-task of mapping bodily activity to subjective experience, which I suggest calling $X$ or chi.

More specifically, recent research has helped explain (modulo $X$) some mental phenomena that had been viewed as mysterious. These include synesthesia, hyper-acuity, subjective contour, word superiority, etc. There is every reason to believe that similar results will be forthcoming in this century, but no way to predict when, if ever, to expect a complete naturalist science of the mind.

What we do know is that the traditional practice of science, taking all relevant phenomena seriously and augmented by ever improving knowledge and techniques, is the only known way forward. It has been especially productive in science to focus effort on issues at the boundary of the known and unknown. The $X$ approach suggests a general way to isolate a core mystery while continuing to study relevant phenomena modulo $X$.

### Acknowledgements

As always, this work benefited from extensive discussions with students and colleagues. David Feldman, who carefully reviewed several drafts, was particularly helpful. Some support was provided by a grant from Google.### References

1. Scientific Realism, Rebecca Goldstein, 2017, in WHAT SCIENTIFIC TERM OR CONCEPT OUGHT TO BE MORE WIDELY KNOWN?, John Brockman Ed, ISBN-10: 0062698214. Edge.org, https://www.edge.org/response-detail/27113

2. The Handbook of Experimental Phenomenology: Visual Perception of Shape, Space and Appearance. Wiley–Blackwell, Albertazzi L., editor (2013) Oxford. ISBN 978-1-119-95468-2